\documentclass{sig-alternate}

\usepackage{graphicx,color}
\usepackage{amsmath,amssymb}
\usepackage{lineno}
\usepackage{algorithm,algorithmic}
\usepackage[sort]{cite}
\usepackage{multicol}
\usepackage{pgfplots}
\usepackage{pgfplotstable}
\usepackage{tikz}
\usetikzlibrary{positioning}
\usetikzlibrary{arrows}
\pgfplotsset{compat=newest}

\def\E{\mathbf{E}}

\let\olditemize\itemize
\renewcommand{\itemize}{
\olditemize
\setlength{\itemsep}{5pt}
\setlength{\parskip}{0pt}
\setlength{\parsep}{0pt}
}

\usepackage{CJKutf8}
\usepackage{ifthen}
\newcommand{\COMM}[2]{{
\begin{CJK}{UTF8}{ipxm}
\ifthenelse{\equal{#1}{TM}}{\color{blue}}{
\ifthenelse{\equal{#1}{HS}}{\color{red}}{
\ifthenelse{\equal{#1}{MI}}{\color{cyan}}{
\ifthenelse{\equal{#1}{CC}}{\color{magenta}}}}}
[#1: #2]
\end{CJK}
}}

\begin{document}

\setcopyright{acmcopyright}
\doi{10.475/123_4}
\isbn{123-4567-24-567/08/06}
\conferenceinfo{WWW '17}{April 3--7, 2017, Perth, Australia}
\acmPrice{\$15.00}

%

\title{Exact Computation of Influence Spread \\ by Binary Decision Diagrams}

\numberofauthors{3} 
\author{
Takanori Maehara$^{1,2)}$ \quad Hirofumi Suzuki$^{3)}$ \quad Masakazu Ishihata$^{3)}$ \\
\affaddr{1) Shizuoka University \ \ 2) RIKEN Center for Advanced Intelligence Project \ \  3) Hokkaido University} \\
\email{takanori.maehara@riken.jp \ \ h-suzuki@ist.hokudai.ac.jp \ \ ishihata.masakazu@ist.hokudai.ac.jp}
}

\maketitle

\begin{abstract}
Evaluating influence spread in social networks is a fundamental procedure to estimate the word-of-mouth effect in viral marketing. 
There are enormous studies about this topic; 
however, under the standard stochastic cascade models, the exact computation of influence spread is known to be \#P-hard.
Thus, the existing studies have used Monte-Carlo simulation-based approximations to avoid exact computation.

We propose the first algorithm to compute influence spread exactly under the independent cascade model.
The algorithm first constructs \emph{binary decision diagrams} (BDDs) for all possible realizations of influence spread,
then computes influence spread by dynamic programming on the constructed BDDs.
To construct the BDDs efficiently, we designed a new \emph{frontier-based search}-type procedure.
The constructed BDDs can also be used to solve other influence-spread related problems, such as random sampling without rejection, conditional influence spread evaluation, dynamic probability update, and gradient computation for probability optimization problems.

We conducted computational experiments to evaluate the proposed algorithm.
The algorithm successfully computed influence spread on real-world networks with a hundred edges in a reasonable time,
which is quite impossible by the naive algorithm.
We also conducted an experiment to evaluate the accuracy of the Monte-Carlo simulation-based approximation by comparing exact influence spread obtained by the proposed algorithm.
\end{abstract}


\newpage

\section{Introduction}
\label{sec:introduction}

\subsection{Background and Motivation}

\emph{Viral marketing} is a strategy to promote products by giving free (or discounted) items to a selected group of highly influential individuals (\emph{seeds}), in the hope that through \emph{word-of-mouth} effects, a significant product adoption will occur~\cite{domingos2001mining,richardson2002mining}.
To maximize the number of adoptions, Kempe, Kleinberg, and Tardos~\cite{kempe2003maximizing} mathematically formalized the dynamics of information propagation, and proposed the optimization problem, referred to as the \emph{influence maximization problem}. 
Several cascade models have been proposed, and the most commonly used one is the \emph{independent cascade model}, proposed by Goldberg, Libai, and Muller~\cite{goldenberg2001talk,goldenberg2001using}.
In this model, the individuals are affected by information
that is stochastically and independently propagated along edges in the network from the seed (Section~\ref{sec:ICmodel}).
To date, significant efforts have been devoted to the development of efficient algorithms for the influence maximization problem~\cite{ohsaka2014fast,cohen2014sketch,borgs2014maximizing,tang2014influence,ohsaka2016dynamic,chen2009efficient,chen2010scalable,cheng2013staticgreedy}.

Here we consider the computational complexity of the influence maximization problem.
Under the independent cascade model, the expected size of influence spread is a non-negative submodular function~\cite{kempe2003maximizing}; thus, a $(1-1/e)$ approximate solution can be obtained by using a greedy algorithm~\cite{nemhauser1978analysis}.
However, the evaluation of influence spread is \#P-hard~\cite{chen2010scalable} because it contains the problem of counting $s$-$t$ connected subgraphs~\cite{valiant1979complexity}.
Thus all existing studies avoided the exact computation 
and employed the Monte-Carlo simulation-based approximation, which simulates the dynamics of information propagation sufficiently many times (e.g., $\Omega(1/\epsilon^2)$) to obtain an accurate (e.g., $1 \pm \epsilon$) approximation of influence spread~\cite{ohsaka2014fast} (Section~\ref{sec:relatedwork}). 

In this study, we first tackle the problem of \emph{computing influence spread exactly under the independent cascade model.}
As the problem is \#P-hard, we are interested in an algorithm that runs on small real-world networks (i.e., having a few hundred edges) in a reasonable time.
The motivations for this studies are as follows.
\begin{itemize}

\item
Influence spread over small networks is practically important.
Because real social networks often consist of many small communities, it is reasonable to consider each community separately or consider only the inter-community network.

\item
When we wish to rank vertices according to their influence spread, we need to compute the values accurately.
Monte-Carlo simulation cannot be used for this purpose 
because it requires $\Omega(1 / \epsilon^2)$ samples for $1 \pm \epsilon$ approximation; 
thus $\epsilon < 10^{-5}$ is impossible.
On the other hand, an exact method can be used 
because its complexity does not depend on the desired accuracy.

\item
Exact influence spread helps to analyze the quality of Monte-Carlo simulation.
Although many experiments using Monte-Carlo simulation have been conducted, none have been compared with the exact value because there is no algorithm that can compute this value.

\item
Establishing a practical algorithm for the fundamental \#P-hard problem is interesting and important task in computer science.
\end{itemize}

\subsection{Contributions}

In this study, we provide the following contributions.

\begin{itemize}
\item 
We propose an algorithm to compute influence spread exactly under the independent cascade model. Note that this is the first attempt to compute this value exactly (Section~\ref{sec:algorithm}). 

\item 
The proposed algorithm enumerates all spread patterns using \emph{binary decision diagrams} (BDDs).
Then, it computes influence spread by dynamic programming on the BDDs.
Here, we have designed a new \emph{frontier-based search} method, which constructs the BDD for $s$-$t$ connected subgraphs efficiently (Section~\ref{sec:frontier}). This is the main technical contribution of this study.

\item
We conducted computational experiments to evaluate the proposed algorithm (Section~\ref{sec:experiments}).
We obtained the exact influence on real-world and synthetic networks with a hundred edges in reasonable times.
We also compared the obtained exact influence with the one obtained using the Monte-Carlo simulation.
\end{itemize}

In addition, using the constructed BDDs, we can also solve the following influence-spread related problems (Section~\ref{sec:application}).
\begin{itemize}
\item 
Random sampling from the set of realizations that successfully propagates information helps to understand the route of influence spread.
We can perform this \emph{without rejection} by using the BDD.

\item
The conditional expectation of the influence spread under the influenced (and non-influenced) conditions on some vertices can be used to measure the effect of conducted viral promotion from a small observations.
This value is efficiently computed by the BDDs.

\item 
When the activation probability changes, we can efficiently update the influence spread. 

\item
The derivatives of the influence spread with respect to the activation probabilities can be computed.
This is used to implement a gradient method for the influence spread optimization problem.
\end{itemize}

\section{Preliminaries}
\label{sec:preliminaries}

\subsection{Independent Cascade Model for Influence Spread}
\label{sec:ICmodel}

The independent cascade model~\cite{goldenberg2001talk,goldenberg2001using} is the most commonly used stochastic cascade model used for social network analysis.
The dynamics of this model is given as follows.

Let $G = (V, E)$ be a directed graph with vertices $V$ and edges $E$. 
Each edge $e \in E$ has \emph{activation probability} $p(e)$.
Each vertex is \emph{either} active or \emph{inactive}.
Note that inactive vertices may become active, but not vice versa.
Here, an active vertex is considered ``influenced.''

Suppose that information is propagated from $S \subseteq V$, 
which is called \emph{seeds}.
Initially, all vertices are inactive.
Then, propagation over the network is performed as follows.
First, each seed $u \in S$ is activated. 
When $u$ first becomes active, it is given a single chance to activate each currently inactive neighbor $v$ with probability $p((u,v))$.
This process is repeated until no further activations are possible.
The expected number of activated vertices after the end of the process is called \emph{influence spread}, which is denoted as $\sigma(S)$.

There is a useful interpretation of influence spread with this model.
We select each edge $e \in E$ with probability $p(e)$. 
Then, we obtain edge set $F$. 
We then consider the induced subgraph $G[F] = (V, F)$, which is a network consisting of only the selected edges.
Here, let $\sigma(S; F)$ be the number of vertices reachable from some $u \in S$ on $G[F]$.
Then, we obtain the following:
\begin{align}
\label{eq:sigma}
	\sigma(S) = \E[ \sigma(S; F) ] = \sum_{F \subseteq E} \sigma(S; F) p(F)
\end{align}
where
\begin{align}
	p(F) = \prod_{e \in F} p(e) \prod_{e' \in E \setminus F} (1 - p(e')).
\end{align}
We use this formula to compute the influence spread.

\subsection{Binary decision diagram}
\label{sec:BDD}

As discussed in Section~\ref{sec:algorithm},
the exact evaluation of \eqref{eq:sigma} involves enumerating $S$-$t$ connecting subgraphs, which is the graph having a path from $S$ to $t$.
To maintain exponentially many such subgraphs,
we use the \emph{binary decision diagram} (BDD),
which is a data structure to represent a Boolean function compactly based on Shannon decomposition.
Note that a Boolean function can be used to represents set family as the indicator function.

A BDD is a directed acyclic graph $D = (\mathcal{N}, \mathcal{A})$ with node set $\mathcal{N}$ and arc set $\mathcal{A}$.%
\footnote{To avoid confusion, we use the terms ``vertex'' and ''edge'' to refer to a vertex and edge in the original graph $G$, 
and ``node'' and ``arc'' to refer to a vertex and edge in the BDD $\mathcal{D}$. 
Vertices are denoted using Roman letters ($u, v, \ldots$) and nodes are denoted using Greek letters ($\alpha, \beta, \ldots$).}
It has two terminals $0$ and $1$.
Each non-terminal node $\alpha \in \mathcal{N}$ is associated with variable $e \in E$, and has two arcs called $0$-arc and $1$-arc.
The nodes pointed by $0$-arc and $1$-arc are referred to as $0$-child and $1$-child (denoted by $\alpha_0$ and $\alpha_1$), respectively.
A BDD represents a Boolean function as follows: A path from the root node to the $1$-terminal represents a (possibly partial) variable assignment for which the represented Boolean function is $\texttt{True}$.
As the path descends to a $0$-arc ($1$-arc) from a node, the node's variable is assigned to $\texttt{False}$ ($\texttt{True}$).

A special type of BDD, i.e., \emph{reduced ordered binary decision diagram} (ROBDD)~\cite{bryant1986graph},
is frequently used in practice.
A BDD is ordered if different variables appear in the same order on all paths from the root. 
A BDD is reduced if the following two rules are applied as long as possible:
\begin{align}
\label{eq:share}
\begin{tabular}{ll}
1. & Share any isomorphic subgraphs. \\
2. & Eliminate all nodes whose two arcs point to \\ & the same node.
\end{tabular}
\end{align}
These rules eliminate redundant nodes in the BDD.
Moreover, when ordering is specified, the ROBDD is determined uniquely~\cite{bryant1986graph}.
In terms of Boolean functions, the function represented by the subgraph rooted by $\alpha$ corresponds to a Shannon co-factor.
The above two rules correspond to sharing nodes with the same Shannon co-factor.
In this paper, we use the term BDD to refer to ROBDD. 

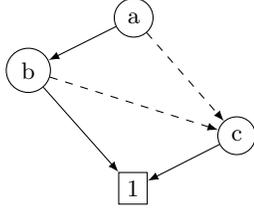
\begin{figure}[tb]
\caption{BDD for $\{ \{c\}, \{a,b\}, \{a,c\}, \{b,c\}, \{a,b,c\} \}$. the $0$-arc is denoted by the dotted line and the $1$-arcs are denoted by the solid lines. The arcs to $0$-terminal are omitted.}
\label{fig:bddexample}
\centering
\begin{tikzpicture}
    \node[circle, draw] (a) {a};
    \node[circle, draw, below left=0.3 and 1 of a] (b) {b};
    \node[circle, draw, below right=1.2 and 1 of a] (c) {c};
    \node[rectangle, draw, below left =0.3 and 1 of c] (1) {1};
    \foreach \u / \v in {a/b, b/1, c/1}
       \draw[-latex] (\u) -- (\v);
    \foreach \u / \v in {a/c,b/c}
      \draw[dashed,-latex] (\u) -- (\v);
\end{tikzpicture}
\end{figure}

Figure~\ref{fig:bddexample} shows an example of BDD,
whihch represents set family $\{\{c\},\{a,b\},\{a,c\},\{b,c\},\{a,b,c\}\}$. The indicator function is 
$\phi(a,b,c) = a (b + \bar b c) + \bar a c$, which corresponds to the diagram.

One important feature of BDD is that it allows efficient manipulation of set families.
In particular, when two set families are represented by BDDs $\mathcal{D}_1$ and $\mathcal{D}_2$ with the same variable ordering, the union and intersection of these BDDs are performed in $O(|\mathcal{D}_1||\mathcal{D}_2|)$ time.
The complement of a set family represented by BDD $\mathcal{D}$ is performed in $O(|\mathcal{D}|)$ time~\cite{bryant1986graph,sieling1993reduction}.
This property is utilized in this study.

For details about BDDs, see the latest volume of ``The Art of Computer Programming''~\cite{knuth2009art} by Knuth.

\section{Algorithm}
\label{sec:algorithm}

In this section, we propose an algorithm to compute influence spread exactly.
Let $S \subseteq V$ be a seed set and $t \in V$ be a vertex.
We consider the set of $S$-$t$ connecting subgraphs
\begin{align}
  \mathcal{R}(S, t) = \{ F \subseteq E : t \text{ is reachable from } S \text{ on } G[F]\},
\end{align}
which represents all realizations in which $t$ is activated from seed set $S$. 
Using this set, influence spread is expressed as
\begin{align}
\label{eq:RIS}
  \sigma(S) = \sum_{t \in V} \sigma(S, t),
\end{align}
where $\sigma(S,t)$ is the influence probability from $S$ to $t$, i.e.,
\begin{align}
	\sigma(S, t) = p(\mathcal{R}(S,t)) = \sum_{F \in \mathcal{R}(S, t)} p(F).
\end{align}
Our algorithm computes influence spread based on the above formulas.
The algorithm first constructs the BDD for $\mathcal{R}(S, t)$.
Then it computes $\sigma(S, t)$ by dynamic programming on the BDD. Finally, by summing over $t \in V$, we obtain the influence spread $\sigma(S)$. 

\subsection{Influence Spread Computation}
\label{sec:DP}

Once BDD $\mathcal{D}(S, t)$ for $\mathcal{R}(S, t)$ is obtained, $\sigma(S, t)$ is efficiently obtained by bottom-up dynamic programming as follows.
Each node $\alpha \in \mathcal{N}$ stores value $\mathcal{B}(\alpha)$, which is the sum of the probabilities of all subsets represented by the descendants of $\alpha$, called the \emph{backward probability}. 
The backward probabilities of $0$-terminal and $1$-terminal are initialized to $\mathcal{B}(0) = 0$ and $\mathcal{B}(1) = 1$.
We process the nodes in reverse topological order (i.e., the terminals to the root).
For each non-terminal node $\alpha \in \mathcal{N} \setminus \{0, 1\}$ associated with edge $e(\alpha) \in E$, $\mathcal{B}(\alpha)$ is computed as follows:
\begin{align}
	\mathcal{B}(\alpha) = (1 - p(e(\alpha))) \mathcal{B}(\alpha_0) + p(e(\alpha)) \mathcal{B}(\alpha_1).
\end{align}
This gives a dynamic programming algorithm (Algorithm~\ref{alg:DP}).
The backward probability of the root node is $\sigma(S, t)$.

\begin{algorithm}[tb]
\caption{Influence spread computation}
\label{alg:DP}
\begin{algorithmic}[1]
\STATE{Create BDD $\mathcal{D} = (\mathcal{N}, \mathcal{A})$ for $\mathcal{R}(S, t)$}
\STATE{Set $\mathcal{B}(0) = 0$, $\mathcal{B}(1) = 1$}
\FOR{$\alpha \in \mathcal{N} \setminus \{0,1\}$ in the reverse topological order}
\STATE{$\mathcal{B}(\alpha) = (1 - p(e(\alpha))) \mathcal{B}(\alpha_0) + p(e(\alpha)) \mathcal{B}(\alpha_1)$}
\ENDFOR
\RETURN{$\mathcal{B}(\text{root})$}
\end{algorithmic}
\end{algorithm}

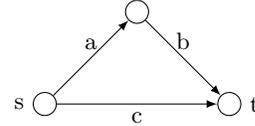
\begin{figure}[tb]
\caption{A graph for example. The BDD for $\mathcal{R}(\{s\}, t)$ is shown in Figure~\ref{fig:bddexample}.}
\label{fig:example}
\centering
\begin{tikzpicture}
    \node[circle, draw] (s) {};
    \node[left=0em of s] {s};
    \node[circle, draw, above right=of s] (u) {};
    \node[circle, draw, below right=of u] (t) {};
    \node[right=0em of t] {t};
    \draw[-latex] (s) -- node[below] {c} ++ (t);
    \draw[-latex] (s) -- node[above] {a} ++ (u);
    \draw[-latex] (u) -- node[above] {b} ++ (t);
\end{tikzpicture}
\end{figure}

Here, we provide an example to illustrate the procedure.
Consider the graph shown in Figure~\ref{fig:example}, which has three edges ($a$, $b$, and $c$).
These activation probabilities are $p$.
Then, the $\{s\}$-$t$ connecting subgraphs are as follows:
\begin{align}
  \mathcal{R}(\{s\},t) = \{\{c\},\{a,b\},\{a,c\},\{b,c\},\{a,b,c\}\}.
\end{align}
The BDD for this set family is presented in Figure~\ref{fig:bddexample}.
We perform dynamic programming on this BDD as follows:
\begin{align*}
\mathcal{B}(1) &= 1, \quad \mathcal{B}(c) = p, \quad \mathcal{B}(b) = p + (1-p) p, \\
\mathcal{B}(a) &= p^2 + (1-p) p^2 + (1-p) p = p + p^2 - p^3 X. 
\end{align*}
Therefore the influence probability from $\{s\}$ to $t$ is $p + p^2 - p^3$.

\subsection{BDD Construction}
\label{sec:frontier}

Here, we present an algorithm to construct the BDD $\mathcal{D}(S, t)$ for $\mathcal{R}(S, t)$.
This is the main technical contribution of this study.

We first consider the single seed case (i.e., $S = \{s\}$) in Section~\ref{sec:single}.
Then, we consider a general case in Section~\ref{sec:multiple}.
For simplicity, we write $\mathcal{R}(s, t)$ and $\mathcal{D}(s, t)$ for $\mathcal{R}(\{s\}, t)$ and $\mathcal{D}(\{s\}, t)$, respectively.

\subsubsection{BDD for a single seed}
\label{sec:single}

Our algorithm is a type of \emph{frontier-based search},
which is a general procedure for enumerating all constrained subgraphs~\cite{kawahara2014frontier}.%
\footnote{Frontier-based search is often applied to construct a \emph{zero-suppressed BDD}, which is a special kind of BDD. However, in our problem, the set has many ``don't care'' edges; therefore BDD is more suitable than ZDD.}
In the following, we first describe the general framework of the frontier-based search.
Then, to adapt it to our problem, we describe four main components: \emph{configuration},
\emph{\texttt{isZeroTerminal} function}, \emph{\texttt{isOneTerminal} function}, and \emph{\texttt{createNode} function}.
Finally we describe two techniques to improve performance: \emph{edge ordering} and \emph{preprocessing}.

\paragraph{Frontier-based search}


Let us enumerate all constrained subgraphs $\mathcal{R} \subseteq 2^E$.
We fix an ordering of edges $(e_1, \ldots, e_m)$ and process the edges one by one, as the exhaustive search.
The processed edges and the unprocessed edges at the end of $i$-th step are denoted by $E^{\le i} := \{ e_1, \ldots, e_i \}$ and $E^{> i} := \{ e_{i+1}, \ldots, e_{m} \}$, respectively.
The set of vertices that has both processed and unprocessed edges is called the \emph{frontier} (at the $i$-th step) and denoted by $W_{i}$.

The set of nodes $\mathcal{N}_i$ represents all subsets of $E^{\le i}$ that can possibly belongs to $\mathcal{R}$.
Each $\alpha \in \mathcal{N}_i$ represents possibly many subsets $R(\alpha) \subseteq 2^{E^{\le i}}$ by paths from the root to $\alpha$,
where a path from the root to $\alpha$ represents a subset in which $e$ is present in the set if the path descends the $1$-arc of node $\beta$ associated with $e$. 
We say that two edge sets $F$ and $F'$ are \emph{equivalent} if for any subsets $H \subseteq E^{> i}$, both $F \cup H$ and $F' \cup H$ belong to $\mathcal{R}$ or neither belong to $\mathcal{R}$.
The algorithm maintains that all sets in $R(\alpha)$ are equivalent.

At the $i$-th iteration, the algorithm constructs $\mathcal{N}_{i}$ from $\mathcal{N}_{i-1}$.
For each node $\alpha \in \mathcal{N}_{i-1}$, the algorithm generates two children for which $e_i$ is excluded or included in the sets in $R(\alpha)$.
Here, the important feature is \emph{node merging}.
Let $\beta$ and $\beta'$ be nodes generated at the $i$-th step.
If all $F \in R(\beta)$ and $F' \in R(\beta')$ are equivalent, we can merge them to reduce the number of nodes.
To verify this equivalence efficiently, each node $\beta$ maintains a data $\phi(\beta)$, referred to as \emph{configuration}, which satisfies the condition that: if $\phi(\beta) = \phi(\beta')$ then the all corresponding sets are equivalent.
Note that the inverse is not required, which causes redundant node expansions.
%

After the process, the constructed BDD is not necessarily reduced.
Thus, we repeatedly apply the reduction rules \eqref{eq:share}.
This reduction is performed in time proportional to the size of the BDD~\cite{bryant1986graph}.

The general framework of the frontier-based search is shown in Algorithm~\ref{alg:frontier},
which contains three auxiliary functions. 
$\texttt{isZeroTerminal}(\alpha, e_i, x)$ ($\texttt{isOneTerminal}(\alpha, e_i, x)$) determines whether the node for the sets excluding (including) $e_i$ from $R(\alpha)$ is the $0$-terminal ($1$-terminal).
More precisely, these are defined as follows:
\begin{align}
  &\texttt{isZeroTerminal}(\alpha, e_i, x) \notag \\
  & = \begin{cases}
  \texttt{True} & \text{all } x\text{-descendants are excluded from } \mathcal{R}, \\
  \texttt{False} & \text{otherwise},
  \end{cases}\\
  &\texttt{isOneTerminal}(\alpha, e_i, x) \notag \\
  &= \begin{cases}
  \texttt{True} & \text{all } x\text{-descendants are included to } \mathcal{R}, \\
  \texttt{False} & \text{otherwise}.
  \end{cases}
\end{align}
$\texttt{createNode}(\alpha, e_i, x)$ creates an $x$-child of $\alpha$. 
To adapt the general framework to our $s$-$t$ connecting subgraph enumeration problem, we only have to design the configuration and these functions.

\begin{algorithm}[tb]
\caption{Frontier-based search}
\label{alg:frontier}
\begin{algorithmic}[1]
\STATE{$\mathcal{N}_0 \leftarrow \{\text{root}\}$, $\mathcal{N}_i \leftarrow \emptyset$ for $i = 1, 2, \ldots, |E|$}
\FOR{$i = 1, 2, \ldots, |E|$}
\FOR{$\alpha \in \mathcal{N}_{i-1}$}
\FOR{$x \in \{0,1\}$}
\IF{$\texttt{isZeroTerminal}(\alpha, e_i, x)$}
\STATE{$\alpha_x \leftarrow 0$}
\ELSIF{$\texttt{isOneTerminal}(\alpha, e_i, x)$}
\STATE{$\alpha_x \leftarrow 1$}
\ELSE
\STATE{$\beta \leftarrow \texttt{createNode}(\alpha, e_i, x)$}
\IF{$\phi(\beta) = \phi(\beta')$ for some $\beta' \in \mathcal{N}_{i}$}
\STATE{$\beta \leftarrow \beta'$}
\ELSE
\STATE{$\mathcal{N}_{i} \leftarrow \mathcal{N}_{i} \cup \{\beta\}$}
\ENDIF
\STATE{$\alpha_x \leftarrow \beta$}
\ENDIF
\ENDFOR
\ENDFOR
\ENDFOR
\STATE{Reduce the constructed BDD by the reduction rules \eqref{eq:share}}
\end{algorithmic}
\end{algorithm}

\paragraph{Configuration}

For two nodes $\beta, \beta' \in \mathcal{N}_{i}$, 
we want to merge these nodes if these are equivalent, i.e., 
the $s$-$t$ reachabilities on $G[F \cup H]$ and $G[F' \cup H]$ are the same for all $F \in R(\beta)$, $F' \in R(\beta')$, and $H \subseteq E^{>i}$.
Thus the configuration must satisfy that $\phi(\beta) = \phi(\beta')$ implies the above condition.

Here, we propose to use the reachability information on the frontier vertices as the configuration as follows.
Let $W_{i}^{s+}, W_{i}^{+t} \subseteq W_i$ be the set of frontier vertices that are reachable from $s$ and reachable to $t$, respectively, on $G[F]$ where $F \in R(\beta)$. 
Note that these are well-defined, i.e., they are independent of the choice of $F$, as mentioned below.
Let $W_{i}^{s-} = W_{i} \setminus W_{i}^{s+}, W_{i}^{-t} = W_{i} \setminus W_{i}^{+t}$.
We define the configuration $\phi(\beta)$ 
as a matrix indexed by $(W_{i}^{s-} \cup \{s\}) \times (W_{i}^{-t} \cup \{t\})$
whose entries denote reachability on $G[F]$: 
\begin{align}
	\phi(\beta)_{uv} = 
    \begin{cases}
    1 & v \text{ is reachable from } u \text{ on } G[F], \\
    0 & \text{otherwise}.
    \end{cases}
\end{align}
If $F \cup H$ admits (does not admit) an $s$-$t$ path, any $F' \in R(\beta')$ with $\phi(\beta) = \phi(\beta')$ also admits (does not admit) an $s$-$t$ path because we can transform the $s$-$t$ path on $G[F]$ to that on $G[F']$ by reconnecting the path on the frontier.
This shows that $\phi$ satisfies the configuration requirement described above.
This also proves, by induction, that this definition is well-defined, i.e., $\phi(\beta)$ is independent of the choice of $F$.

\paragraph{``isZeroTerminal'' and ''isOneTerminal'' functions}

If $x = 1$, i.e., we include edge $e_i = (u, v)$ in the sets in $R(\alpha)$, we have a chance to obtain $\texttt{isOneTerminal}(\alpha, e_i, x) = \texttt{True}$,
which is the case that the included edges contain a path from $s$ to $t$.
Using our configuration, this is easily implemented as follows:
\begin{align}
	&\texttt{isOneTerminal}(\alpha, e_i, 1) \notag \\
    & \qquad = 
    \begin{cases}
    	\texttt{True} & \phi(\alpha)_{su} = 1 \text{ and } \phi(\alpha)_{vt} = 1, \\
        \texttt{False} & \text{otherwise}.
    \end{cases}
\end{align}
Similarly, if $x = 0$, i.e., we exclude edge $e_i$ from the sets in $R(\alpha)$, we have a chance to obtain $\texttt{isZeroTerminal}(\alpha, e_i, x) = \texttt{True}$, which is the case that the excluded edges form a cutset from $s$ to $t$.
This is implemented as follows.
\begin{align}
	&\texttt{isZeroTerminal}(\alpha, e_i, 0) \notag \\
    & = 
    \begin{cases}
    	\texttt{True} & t \text{ is unreachable from } s \text{ on } G[F \cup E^{>i}], \\
        \texttt{False} & \text{otherwise}
    \end{cases}
\end{align}
where $F \in R(\alpha)$.
Note that this is well-defined for the same reason described above.
To check the reachability on $G[F \cup E^{>i}]$ efficiently, we precompute the transitive closures of $G[E^{>j}]$ for all $j = 0, 1, \ldots, |E|$.%
\footnote{Because we compute the BDDs for all pairs of $s, t \in V$, storing all transitive closures accelerates computation. The size of all transitive closures are typically much smaller than the size of the BDDs.}
Then the reachability from $s$ to $t$ is checked in $O(|W_i|^2)$ time by the DFS/BFS with the configuration and the precomputed reachability.

\paragraph{``createNode'' function}

The most important role of $\texttt{createNode}(\alpha, e_i, x)$ is computing the configuration of the new node.
The function first creates new node $\beta$ and copies configuration $\phi(\alpha)$ to $\phi(\beta)$.
If a vertex is included in the frontier (i.e., some incident edge is processed first) or excluded from the frontier (i.e., all incident edges have been processed), 
we insert or remove the corresponding row and column from the configuration $\phi(\beta)$.

If $x = 0$, we require no further updates.
Otherwise, adding a new edge changes reachability;
thus we update $\phi(\beta)$ to be the transitive closure of the frontier.
This is performed in $O(|W_i|^2)$ time by the DFS/BFS on the frontier.

\paragraph{Edge ordering}

The complexity of the frontier-based search depends on the frontier size.
$\mathcal{N}_{i}$ has at most $O(2^{|W_{i}|^2})$ nodes
because it contains no nodes with the same configurations.
It is known that the frontier size is closely related to the \emph{pathwidth} graph parameter~\cite{kinnersley1992vertex}.

Note that optimizing edge ordering is important to reduce the frontier size (i.e., the pathwidth).
For our problem, there is an additional requirement, i.e.,
the same edge ordering is used for all BDDs $\mathcal{R}(s, t)$ for $s, t \in V$
because we perform several set manipulations between the BDDs.

In this study, we use the \emph{path-decomposition based ordering} proposed by Inoue and Minato~\cite{inoue2016acceleration}.
The algorithm first computes a path decomposition with a small pathwidth using beam search-based heuristics. 
Then it computes an edge ordering using the path decomposition information.

\paragraph{Preprocessing}

If $e \in E$ is not contained in any $s$-$t$ simple path, $e$ does not appear in the BDD because the existence of $e$ does not affect $s$-$t$ reachability.
Therefore, removing all such edges as a preprocessing improves the performance of the algorithm.

Determining whether there is an $s$-$t$ simple path containing $e$ is NP-hard because it reduces to the NP-hard two-commodity flow problem~\cite{even1975complexity}.
However, because we are interested in small networks, we can enumerate all $s$-$t$ simple paths using Knuth's Simpath algorithm~\cite{knuth2009art},
which is a frontier-based search algorithm that runs faster than the proposed algorithm because it uses a smaller configuration.
Thus, we can use the Simpath algorithm in preprocessing.

\subsubsection{BDD for multiple seeds}
\label{sec:multiple}

The frontier-based search described in the previous subsection can be easily adopted to the multiple seeds case.
However, there is a more efficient way to construct the BDD for multiple seeds. 

The method is based on the following formula, which is immediately obtained from the definition of $\mathcal{R}(S, t)$:
\begin{align}
\mathcal{R}(S, t) = \bigcup_{s \in S} \mathcal{R}(s, t).
\end{align}
Because the BDD of the union of two set families represented by BDDs $\mathcal{D}_1$ and $\mathcal{D}_2$ is obtained in $O(|\mathcal{D}_1| |\mathcal{D}_2|)$ time, 
and, practically, the size of the BDDs is small (Section~\ref{sec:experiments}),
this approach is more efficient than the frontier-based  practice.


%
%

\subsubsection{Node sharing among BDDs}

To compute influence spread, we construct BDDs for all pairs of $s, t \in V$. 
Here, intuitively, if two source-target pairs $(s,t)$ and $(s',t')$ are close, the BDDs $\mathcal{D}(s,t)$ and $\mathcal{D}(s',t')$ may share many subgraphs. 
Thus, by sharing the nodes corresponding to the subgraphs, we can reduce the total size of the BDDs~\cite{minato1990shared}.
This also reduces the total complexity of computing influence spreads for all source-target pairs $(s,t)$, which is proportional to the total size of the shared BDDs.

\section{Other Applications}
\label{sec:application}

In the previous section, we established an algorithm to construct the BDD for all $S$-$t$ connecting subgraphs $\mathcal{R}(S, t)$.
This data structure allows us to solve influence spread-related problems efficiently.

\subsection{Random Sampling without Rejection}
\label{sec:sampling}

Sometimes we want to know how the influence is propagated from $S$ to $t$.
The random sampling from $\mathcal{R}(S, t)$ will help us to understand this; however, the naive method that performs Monte-Carlo simulation and rejects if $S$ does not connect to $t$ usually requires impractically many simulations due to the small influence probability.
Here we show that this random sampling can be performed \emph{without rejection} using BDD $\mathcal{D}(S, t) = (\mathcal{N}, \mathcal{A})$~\cite{ishihata2011bayesian}.

As a preprocess, we perform the dynamic programming described in Section~\ref{sec:DP} to compute the backward probability $\mathcal{B}(\alpha)$ for each node $\alpha \in \mathcal{N}$.
Then, we perform the following random walk, which starts from the root node and ends at the $1$-terminal:
When we are on non-terminal node $\alpha \in \mathcal{N} \setminus \{0, 1\}$ associated with $e \in E$,
we randomly move $\alpha_0$ or $\alpha_1$ with probability proportional to $(1 - p(e)) \mathcal{B}(\alpha_0)$ and $p(e) \mathcal{B}(\alpha_1)$.
Here, if we moved to $\alpha_0$, we exclude $e$ from $F$; otherwise we include $e$ in $F$.
We repeat this procedure until we reach the $1$-terminal.
Finally, for all undetermined edges, we randomly and independently exclude or include the edge with its probability.
This yields a random sampling from $\mathcal{R}$. The complexity is proportional to the height of the BDD.

\subsection{Conditional Influence Spread}

After conducting a viral promotion, we must measure the effect of the promotion.
For this purpose, we observe the status of influence (i.e., influenced or not) on some small vertices
and estimate the total size of influence spread.
This value, referred to as the \emph{conditional influence spread}, can be 
obtained using the constructed BDDs.

For example, suppose that we have observed that ``vertices $u, v$ are influenced and $w$ is not influenced.'' 
Then, the realizations that satisfy this condition is given by
\begin{align}
	\mathcal{R} = \mathcal{R}(S, u) \cap \mathcal{R}(S, v) \cap \mathcal{R}(S, w)^c,
\end{align}
where $\mathcal{R}(S, w)^c = 2^E \setminus \mathcal{R}(S, w)$. 
Then the conditional influence probability from $S$ to $t$ under $\mathcal{R}$ is given by
\begin{align}
\label{eq:conditional}
  \sigma(S, t | \mathcal{R}) = \frac{ p(\mathcal{R}(S, t) \cap \mathcal{R}) }{ p(\mathcal{R}) },
\end{align}
and the summation over $t$ gives the conditional influence spread.

The BDDs for $\mathcal{R}(S,t) \cap \mathcal{R}$ and $\mathcal{R}$ in \eqref{eq:conditional} can be efficiently obtained because Boolean operations on set families are performed efficiently on BDD representations.
Moreover, these probabilities can be computed by the 
the dynamic programming described in Section~\ref{sec:DP}.
This is the method for computing the exact conditional influence spread.

Note that, by combining random sampling technique described in Section~\ref{sec:sampling}, we can sample conditional realizations without rejection.

\subsection{Activation Probability Modification}

Activation probabilities are frequently changed in real-world networks~\cite{ohsaka2016dynamic}.
In such a case, we can recompute the influence spread easily by reusing the constructed BDDs. The complexity is proportional to the size of the BDDs.

\subsection{Activation Probability Optimization}

Sometimes we want to solve an optimization problem with respect to the activation probabilities of edges.
One example is a time-dependent influence problem, i.e., when the activation probabilities are the function on time, we want to seek the time that maximizes influence spread.
Another example is a network design problem where we want to maximize the influence spread by modifying activation probabilities under some (e.g., budget) constraint. 
Because these problems are non-convex optimization problems (even if the activation probabilities are simple functions), it is difficult to compute the optimal solution.
However, a local optimal solution would be obtained by a gradient-based method.

To implement a gradient-based method, we require derivatives of the influence spread with respect to the activation probabilities.
Here we show that if we have the BDD for $\mathcal{R}(S, t)$, we can obtain $\partial \sigma(S,t) / \partial p(e)$ for all $e \in E$ in time proportional to the size of the BDD.

First, we compute the backward probability $\mathcal{B}(\alpha)$ for all nodes $\alpha \in \mathcal{N}$ by the dynamic programming described in Section~\ref{sec:DP}.
Then, we perform top-down dynamic programming as follows.
Each node $\alpha \in \mathbb{N}$ has a value $\mathcal{F}$, called the \emph{forward probability}.
The forward probability of the root node is initialized as $\mathcal{F}(\text{root}) = 1$.
We process the nodes in topological order (i.e., the root to the terminals).
When we are on non-root node $\alpha \in \mathcal{N} \setminus \{\text{root}\}$, its forward probability is determined as follows:
\begin{align}
  \mathcal{F}(\alpha) = \sum_{\beta: \beta_0 = \alpha} (1 - p(e(\beta))) \mathcal{F}(\beta) + \sum_{\gamma: \gamma_1 = \alpha} p(e(\gamma)) \mathcal{F}(\gamma).
\end{align}
Then, the derivative is obtained as follows:
\begin{align}
  \frac{\partial \sigma(S,t)}{\partial p(e)} = \sum_{\alpha: e(\alpha) = e} \mathcal{F}(\alpha) \mathcal{B}(\alpha_1).
\end{align}
Because Monte-Carlo simulation cannot be used to compute the derivative, this is an advantage of our method.
Note that this technique is used in probabilistic logic learning~\cite{ishihata2008propositionalizing,inoue2009evaluating}.

\section{Experiments}
\label{sec:experiments}

We conducted computational experiments to evaluate the proposed algorithm.
All code was implemented in C++ (g++5.4.0 with the -O3 option)
using the \texttt{TdZdd} library\footnote{https://github.com/kunisura/TdZdd}, which is a highly optimized implementation for BDDs.
All experiments were conducted on 64-bit Ubuntu 16.04 LTS with an Intel Core i7-3930K 3.2~GHz CPU and 64~GB RAM.

The real-world networks were taken from the Koblenz Network Collection.\footnote{http://konect.uni-koblenz.de/}
All self-loops and multiple edges were removed,
and undirected edges were replaced with two directed edges in both directions.
The number of vertices and edges are described in Table~\ref{tbl:exp1}

\subsection{Scalability on Real-World Networks}

First, to evaluate the performance of the proposed algorithm in the real-world networks, we conducted experiments on the collected networks. 
For each network, we constructed the BDDs for all distinct $s, t \in V$ and observed the computational time, the size of each BDD, the total shared size of the BDDs, and the number of realizations that are represented by the BDDs (i.e., cardinality of the set).

The results are shown in Table~\ref{tbl:exp1}.
The algorithm successfully computed the BDDs for networks with a hundred edges, but failed on some larger networks.
When it succeeded, it is very efficient in both time and space, i.e., it ran in a few milliseconds and the size was at most a few millions for a network with a few hundred edges.
The shared size was about the half of the sum of all sizes of BDDs, which means that the BDDs shared many nodes.
By comparing Contiguous-USA network and the three failed networks, the computational cost depended on the network structure.
 
It should be emphasized that the naive exhaustive search is quite impractical for these networks
because, as shown in Cardinality column, there are enormous number of connecting realizations.
In particular, at the extreme case, a BDD $\mathcal{D}(s,t)$ for American-Revolution network with some source-target pair $(s,t)$ consisted of only 85 nodes, but represented
\begin{align}
\begin{tabular}{r}
2,058,334,714,926,419,025,286,040,286,320,\\
632,494,993,236,943,086,975,345,403,704,463,\\
133,047,043,046,026,363,318,022,843,662,336\phantom{,}
\end{tabular}
\end{align}
realizations (approximately $2 \times 10^{97}$), which exceeds the number of atoms in the universe (approximately $10^{80}$).
This shows the effectiveness of the BDD representation of the connecting realizations.

\begin{table*}[tb]
\caption{Computational results on real-world networks. Time denotes the average time to construct the BDDs, BDD Size denotes the average number of nodes in the BDDs, Shared Size denotes the total number of distinct nodes in the shared BDDs, and Cardinality denotes the average number of subgraphs represented by the BDDs. Here, average is taken of all distinct $s, t \in V$. For the last three networks, the algorithm failed to compute due to the memory limit.}
\label{tbl:exp1}
\centering
\begin{tabular}{|rrrrrrr|}
\hline
\multicolumn{1}{|c}{Network} & \multicolumn{1}{c}{Vertices} & \multicolumn{1}{c}{Edges} & \multicolumn{1}{c}{Time [ms]} & \multicolumn{1}{c}{BDD Size} & \multicolumn{1}{c}{Shared Size} & \multicolumn{1}{c|}{Cardinality} \\
\hline
South-African-Companies & 11 & 26  & 0.1 & 12.1 & 472 & 2.2e+07 \\
Southern-women-2 & 20 & 28  & 0.3 & 54.7 & 2,266 & 1.3e+08 \\
Taro-exchange & 22 & 78  & 4.1 & 1,119.2 & 277,756 & 1.6e+23 \\
Zachary-karate-club & 34 & 156  & 24.9 & 7,321.8 & 4,988,148 & 6.4e+46 \\
Contiguous-USA & 49 & 214 & 117.9 & 30,599.8 & 41,261,047 & 1.6e+64 \\
American-Revolution & 141 & 320 & 2.2 & 120.0 & 1,530,677 & 5.7e+95 \\
\hline
Southern-women-1 & 50 & 178 & --- & --- & --- & --- \\
Club-membership & 65 & 190 & --- & --- & --- & --- \\
Corporate-Leadership & 64 & 198 & --- & --- & --- & --- \\
\hline
\end{tabular}
\end{table*}

\subsection{Scalability on Synthetic Networks}

Next, to observe the performance of the algorithm precisely, 
we conducted experiment on two classes of synthetic networks.
The first class was $5 \times w$ grid graph, which has $n = 5 w$ vertices and $9w - 5$ undirected edges, which has a pathwidth of $5$.
The second class was the random graph that has the same number of vertices and edges as the grid graph, which has a pathwidth of $\Theta(n)$.
We computed influence probability $\sigma(s,t)$ from the north-west corner $s$ to the south-east corner $t$ on the grid graph and the corresponding vertices on the random graph.

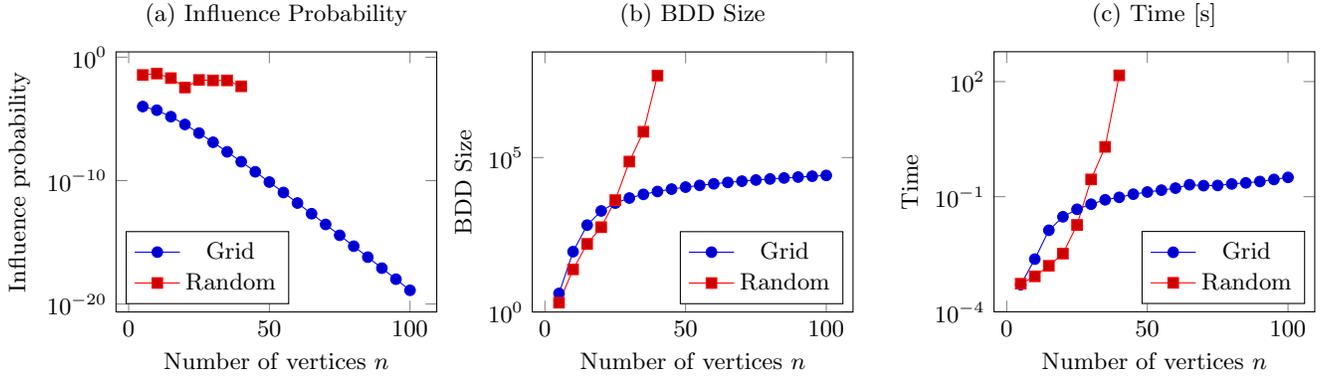
\begin{figure*}[tb]
\begin{minipage}{.33\textwidth}
\begin{tikzpicture}
\begin{semilogyaxis}
[
	scale=1.0,
	xlabel={Number of vertices $n$},
    ylabel={Influence probability},
    title={(a) Influence Probability},
    ylabel near ticks,
    legend pos=south west,
    width=\textwidth
]
\addplot table [x=n, y=inf] {results/exp2g.txt};
\addlegendentry{Grid}
\addplot table [x=n, y=inf] {results/exp2r.txt};
\addlegendentry{Random}
\end{semilogyaxis}
\end{tikzpicture}
\end{minipage}
\begin{minipage}{.33\textwidth}
\begin{tikzpicture}
\begin{semilogyaxis}
[
	scale=1.0,
	xlabel={Number of vertices $n$},
    ylabel={BDD Size},
    title={(b) BDD Size},
    ymin=1,
    ylabel near ticks,
	legend pos=south east,
	width=\textwidth
]
\addplot table [x=n, y=size] {results/exp2g.txt};
\addlegendentry{Grid}
\addplot table [x=n, y=size] {results/exp2r.txt};
\addlegendentry{Random}
\end{semilogyaxis}
\end{tikzpicture}
\end{minipage}
\begin{minipage}{.33\textwidth}
\begin{tikzpicture}
\begin{semilogyaxis}
[
	scale=1.0,
	xlabel={Number of vertices $n$},
    ylabel={Time},
    title={(c) Time [s]},
    ymin=0.0001,
    ylabel near ticks,
    legend pos=south east,
	width=\textwidth
]
\addplot table [x=n, y=time] {results/exp2g.txt};
\addlegendentry{Grid}
\addplot table [x=n, y=time] {results/exp2r.txt};
\addlegendentry{Random}
\end{semilogyaxis}
\end{tikzpicture}
\end{minipage}
\caption{Computational results on $5 \times w$ grid graphs and random graphs. The algorithm failed to compute the influence spread on the random network with $n \ge 45$ vertices due to the memory limit.}
\label{fig:synthetic}
\end{figure*}

The results are shown in Figure~\ref{fig:synthetic}. 
For the grid graphs, BDD size and construction time increased slowly; thus the computation on $n = 100$ was tractable. 
On the other hand, for the random graphs, BDD size and construction time increased rapidly; thus we could not compute a BDD for $n \ge 45$.
These results are consistent with the pathwidths of these networks.

For both networks, the influence probabilities decayed exponentially. 
It decayed faster in grid network since basically the influence probability depends on the network distance.

\subsection{Influence Maximization Problem}
\label{sec:expinfmax}

Here, we consider the influence maximization problem, which seeks $k$ seeds to maximize the influence spread~\cite{kempe2003maximizing}.
The greedy algorithm is commonly used to solve this problem,
which begins from the empty set $S = \emptyset$ and 
repeatedly adds the vertex $u$ that has the maximum marginal influence $\sigma(S \cup \{u\}) - \sigma(S)$ into $S$ until $k$ vertices are added. 

We implemented the greedy algorithm with the exact influence spread to observe the performance of the proposed algorithm in the greedy algorithm.
We used the Contiguous USA network and Zachary Karate club networks. 

\begin{figure*}[tb]
\begin{minipage}{.33\textwidth}
\begin{tikzpicture}
\begin{axis}
[
	scale=1.0,
	ylabel={Influence spread},
    xlabel={Number of steps},
    title={(a) Influence spread},
    ylabel near ticks,
    ymin=0,
    legend pos=south east,
	width=\textwidth
]
\addplot table [x=step, y=total_influence, col sep=comma] {results/exp3usa.txt};
\addlegendentry{ContUSA}
\addplot table [x=step, y=total_influence, col sep=comma] {results/exp3zac.txt};
\addlegendentry{Zachary}
\end{axis}
\end{tikzpicture}
\end{minipage}
\begin{minipage}{.33\textwidth}
\begin{tikzpicture}
\begin{semilogyaxis}
[
	scale=1.0,
	ylabel={Shared Size},
    xlabel={Number of steps},
    title={(b) Shared Size},
    ylabel near ticks,
    ymin=100000,
	legend pos = south east,
	width=\textwidth
]
\addplot table [x=step, y=shared_bdd_size, col sep=comma] {results/exp3usa.txt};
\addlegendentry{ContUSA}
\addplot table [x=step, y=shared_bdd_size, col sep=comma] {results/exp3zac.txt};
\addlegendentry{Zachary}
\end{semilogyaxis}
\end{tikzpicture}
\end{minipage}
\begin{minipage}{.33\textwidth}
\begin{tikzpicture}
\begin{semilogyaxis}
[
	scale=1.0,
	ylabel={Time [s]},
    xlabel={Number of steps},
    title={(c) Time [s]},
    ymin=0.1,
	legend pos=south east,
	width=\textwidth
]
\addplot table [x=step, y=total_time, col sep=comma] {results/exp3usa.txt};
\addlegendentry{ContUSA}
\addplot table [x=step, y=total_time, col sep=comma] {results/exp3zac.txt};
\addlegendentry{Zachary}
\end{semilogyaxis}
\end{tikzpicture}
\end{minipage}
\caption{Computational results on the influence maximization problem with exact influence spread.}
\label{fig:exp3}
\end{figure*}
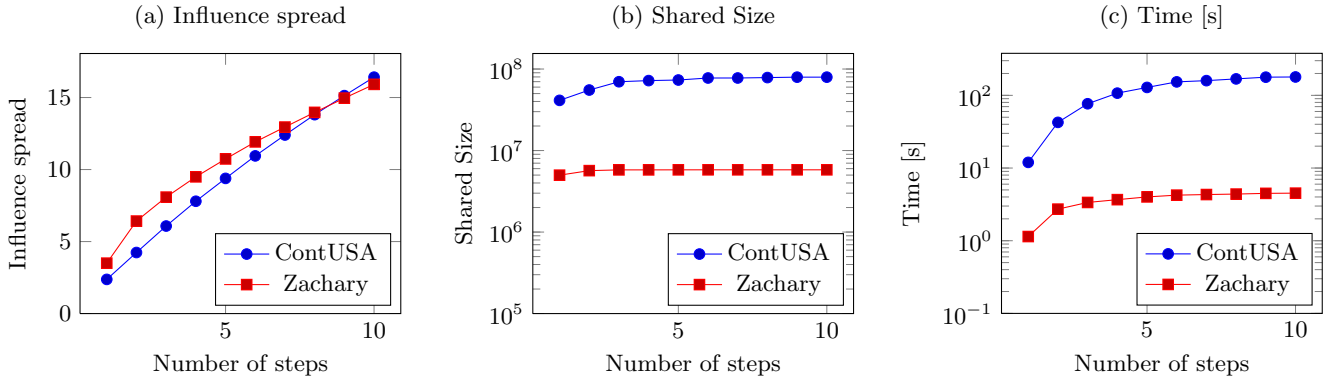

The results are shown in Figure~\ref{fig:exp3}.
Figure~\ref{fig:exp3}(b) shows that the shared size of BDD did not increase while the algorithm process. The shared size at the $10$-th step of the greedy algorithm was two times larger than the $1$-st step for both networks, 
and the computational times were proportional to the number of steps, as shown in Figure~\ref{fig:exp3}(c).

\subsection{Comparison with Monte-Carlo simulation}

Finally, for an application of exact influence spread computation,
we compared the exact influence spread with the Monte-Carlo simulation. 
We used the Contiguous USA network, which was also used in the above experiment.
In addition, we used a seed set of size $10$ computed by the greedy algorithm with the exact influence spread, and we compared the quality of the approximated spread.

\begin{figure}[tb]
\begin{tikzpicture}
\begin{axis}
[
	scale=0.8,
	ylabel={Error of the estimated influence},
    xlabel={Number of samples},
    ymin=-0.002,
    ymax=+0.0015,
    xmin=0,
    ylabel near ticks,
    yticklabel style={
    	/pgf/number format/sci,
    },
]
\addplot table [mark=none,x=samples, y expr={\thisrowno{1} - 16.4145792376}] {results/exp4.txt};
\addplot[domain=0:10000000, samples=100]{0};
\end{axis}
\end{tikzpicture}
\caption{Accuracy of Monte-Carlo simulation.}
\label{fig:exp4}
\end{figure}
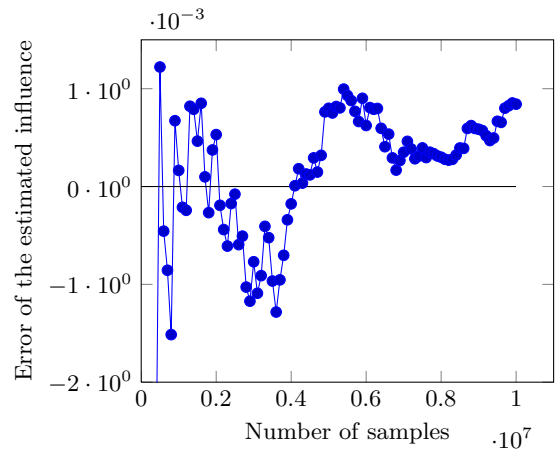

The results are shown in Figure~\ref{fig:exp4}. 
Even for such small size network ($m = 107$ edges) and the large number of Monte-Carlo samples ($N = 10^7$), the estimated influence spread by Monte-Carlo simulation has error in the order of $10^{-3}$, which is consistent with the theory~\cite{ohsaka2014fast}.

\section{Related Work}
\label{sec:relatedwork}

\paragraph{Influence spread computation}

After the seminal work by Kempe, Kleinberg, and Tardos~\cite{kempe2003maximizing}, influence spread over networks has become an important topic in social network analysis.
However, to the best of our knowledge,
no efforts have been devoted to the exact computation of influence spread since it is proved to be \#P-hard by Chen et al.~\cite{chen2010scalable}.

To compute influence spread, all existing studies used Monte-Carlo simulation-based approximation,
which repeats simulation until a reliable estimation is obtained.
This approach is originally proposed in \cite{kempe2003maximizing}. 
To enhance the scalability, many techniques,
such as pruning~\cite{leskovec2007cost} and sample average approximation~\cite{chen2010scalable,cheng2013staticgreedy,ohsaka2014fast} have been investigated.

The recent approximation methods are based on the Borgs et al.'s \emph{reverse influence sampling} (RIS) technique~\cite{borgs2014maximizing},
which randomly selects a vertex and then performs reverse BFS to compute the set of vertices reachable to the selected vertex on a random graph. 
It is important that this procedure is implemented in time proportional to the size of the sample. 
Therefore it successfully bounds the complexity of influence spread approximation.
Tang, Shi, and Xiao~\cite{tang2014influence,tang2015influence} proposed the methods to reduce the number of samples. 

Note that our formulation~\eqref{eq:RIS} is related with the RIS technique: RIS randomly selects vertices whereas we select all vertices, and RIS samples single reverse influence patterns whereas ours enumerates all reverse influence patterns.

\paragraph{Subgraph enumeration}

In this study, 
we virtually solved the enumeration problem of $s$-$t$ connecting subgraphs for the influence spread computation.
This problem is known to be \#P-hard~\cite{valiant1979complexity}.

If the underlying network is undirected, this problem coincides with the \emph{two-terminal network reliability problem}~\cite{valiant1979complexity,brecht1988lower},
and several algorithms have been proposed to construct a BDD for the problem~\cite{hardy2005computing,yan2015novel}.
However, none have been naturally generalized to our directed problem because they essentially exploit the undirected nature of the graph.

BDD is used to enumerate several kinds of subgraphs (substructures), such as paths~\cite{knuth2009art}, spanning trees~\cite{sekine1995computing}, and the solutions of logic puzzles~\cite{yoshinaka2012finding}.
By comparing these methods, the proposed method involves relatively expensive operations (reachability computation) in the auxiliary functions used in the frontier-based search. 
Such operations usually make the algorithm non-scalable; thus these are not used in literature.
However, in our case, these are necessary to scale up the algorithm by pruning many nodes in each step.

\section{Conclusion}
\label{sec:conclusion}

In this study, we have proposed an algorithm to compute influence spread exactly.
The proposed algorithm first constructs the BDDs to represent all $s$-$t$ connecting subgraphs.
Then it computes influence spread by dynamic programming on the constructed BDDs.
The BDDs can also be used to solve some other influence-spread related problems efficiently.
The results of our computational experiments show that the proposed algorithm scales up to networks with a hundred edges,
even though they have an enormous number (i.e., $\sim 2 \times 10^{97}$) of possible realizations.

A similar approach will be adopted for the \emph{linear threshold model}~\cite{kempe2003maximizing}, which is another widely used stochastic cascade model:
Goyal, Lu, and Lakshmanan~\cite{goyal2011simpath} showed that the influence spread in this model is computed by enumerating all $s$-$t$ paths, and they proposed an algorithm, named ``Simpath,'' based on an exhaustive search with pruning.
By constructing the BDDs for all $s$-$t$ paths, rather than for all $s$-$t$ connected subgraphs as in this study, similar results will be obtained.
Note that there is an efficient algorithm to construct the BDD for all $s$-$t$ paths~\cite{knuth2009art}, which is also named ``Simpath.''
This algorithm is used in this study to prune the redundant edges in preprocessing.

The most important future work is computing exact (or highly accurate) influence spread in networks with a few hundred edges or a thousand edges. 
This may require new technique such as parallel construction of BDDs, approximation of BDDs, or exploiting network structures.
\subsection*{Acknowledgment}

\sloppy

This work was supported by JSPS KAKENHI Grant Numbers 15H05711 and 16K16011,
and by JST, ERATO, Kawarabayashi Large Graph Project.


\bibliographystyle{abbrv}
\bibliography{main.bib}

\end{document}